\documentclass[aps,prl,twocolumn,groupedaddress]{revtex4}

\usepackage{amsmath,amsthm,amsfonts, latexsym}

\newcommand{\mtx}[2]{\left(\begin{array}{#1}#2\end{array}\right)}
\begin{document}

%\preprint{}

\title{The rebit three-tangle and its relation to two-qubit entanglement}


\author{William K. Wootters}
%\thanks{}
\affiliation{Department of Physics, Williams College, Williamstown, MA 01267, USA}


%\date{February 6, 2014}

\begin{abstract}
The three-tangle is a measure of three-way entanglement in a system of three qubits.
For a pure state, it can be understood as the residual entanglement not accounted for 
by pairwise entanglements between individual qubits.  Here we define and evaluate 
the analogous quantity for three {\em rebits} (that is, binary systems in the real-amplitude variant of quantum theory).
We find that the resulting formula is the same as in the complex case, {\em except} that an 
overall absolute value sign is missing.  As a result, the rebit three-tangle can be negative,
expressing the possibility of non-monogamous entanglement in real-amplitude quantum theory (for entanglement
based on the convex-roof construction).
We then relate the entanglement among three rebits to the entanglement of two qubits,
by re-expressing the two-qubit state as a three-rebit state in the ubit model.  
\end{abstract}

%\keywords{}

\maketitle


\section {Introduction}

In the classic 1936 paper by Birkhoff and von Neumann on the logical structure of quantum theory, the authors note that 
the logical principles laid out in the paper are satisfied not only by standard quantum theory but also by the analogous
theories in which the usual vector space over the complex field is replaced by a vector space over either the reals or the
quaternions  \cite{Birkhoff} (see also Refs.~\cite{Jordan, Soler, Holland}).  Nature seems to have chosen the complex version, but the other two theories are still of 
interest \cite{Stueckelberg, Finkelstein, Adler}.  Studying these other
theories sheds light on standard quantum theory through the comparison, and it is conceivable that one or the other of these
alternative theories will someday be needed in our account of the physical world.  The real-vector-space version is interesting
also in that it can be used as a framework in which to embed the complex theory, thus providing an alternative mathematical
representation that can be useful even if the physical content does not depart at all from the standard theory \cite{Stueckelberg,
Dyson, Hestenes, Baez}.  
%Both Stueckelberg and Dyson
%have noted, for example, that the time-reversal operation becomes linear, as opposed to antilinear, when quantum theory is re-expressed
%in a real Hilbert space.  

The present paper focuses on the real-vector-space variant of quantum theory, within which we investigate the entanglement of simple 
systems.  Bipartite entanglement in real-amplitude quantum theory has been studied before, for example in Refs.~\cite{Rungta, Batle, Slater,
Shuddhodan}, and  
qualitative work on multi-partite real-amplitude systems has been done in Ref.~\cite{Wootters2}.  
Here we study quantitatively the entanglement in a tripartite system consisting of three rebits, that is, binary systems in the real-vector-space theory.  
For a pure state of three {\em qubits}---call them $\mathcal A$, $\mathcal B$, and $\mathcal C$---there is a well-known relation among various entanglements within the system, expressed in terms
of the bipartite tangle $\tau$ and the three-tangle $\tau_{\mathcal A \mathcal B \mathcal C}$ (see the following section for the definitions):
\begin{equation}  \label{basictau}
\tau_{\mathcal A|\mathcal B\mathcal C} = \tau_{\mathcal A|\mathcal B} + \tau_{\mathcal A|\mathcal C} + \tau_{\mathcal A\mathcal B\mathcal C}.
\end{equation}
That is, as measured by the tangle, 
the entanglement between qubit $\mathcal A$ and the pair $\mathcal B\mathcal C$ can be divided into three parts: (i) the entanglement between
$\mathcal A$ and $\mathcal B$, (ii) the entanglement between $\mathcal A$ and $\mathcal C$, and (iii) the three-way entanglement among all three qubits as expressed
by the three-tangle \cite{Coffman}.  It is particularly interesting that even though Eq.~(\ref{basictau}) singles out qubit $\mathcal A$ as the special ``hinge'' 
for evaluating the tangles, the three-tangle itself is invariant under permutations of the qubits.  In this paper we work out
the analog of Eq.~(\ref{basictau}) for the real-vector-space theory.  We find that the analog of the three-tangle is again invariant under permutations of the elementary subsystems.  Moreover, the 
analytic formula
for the ``rebit three-tangle'' is almost identical to the one for the original three-tangle: the only difference is the absence of
an overall absolute value sign.  It turns out, in fact, that unlike the usual three-tangle, the rebit three-tangle can be negative.
As we will see, the possibility of negative values reflects the possibility of non-monogamous entanglement in the real-vector-space 
theory \cite{Wootters2}.  

Once we have derived our formula for the rebit three-tangle, we use that result to gain a new perspective on the ordinary entanglement between
two qubits.  Here we make use of the ``ubit'' model, in which any state of an ordinary quantum system ${\mathcal S}$ can be 
represented as the state of a corresponding real-amplitude system $S$ together with an auxiliary rebit $U$ called the 
universal rebit or ubit \cite{Aleksandrova}.  (Throughout this paper I use script letters for ordinary quantum systems and plain letters for real-vector-space systems.)
Specifically, we rewrite any state of a pair of qubits ${\mathcal A}{\mathcal B}$ as a state of three rebits $UAB$.  We find that for pure states
of $\mathcal A \mathcal B$, the tangle
between ${\mathcal A}$ and ${\mathcal B}$ is simply the tangle between the corresponding rebits $A$ and $B$, plus the rebit three-tangle of the system $UAB$.  We also show how this result can be extended to mixed states of two qubits.

\section{Review of the three-tangle for qubits}

Though the concept of tangle (or the closely related concept of concurrence) has been defined for quantum systems of arbitrary dimension \cite{Caves, Osborne} and with arbitrarily 
many subsystems \cite{Mintert1, Mintert2}, here we restrict our attention to the simple case of two qubits.  
For a pure state $|\psi\rangle$ of two qubits ${\mathcal A}$ and ${\mathcal B}$, $|\psi\rangle = a|00\rangle + b|01\rangle + c|10\rangle +d|11\rangle$, the tangle
between ${\mathcal A}$ and ${\mathcal B}$ is defined to be 
\begin{equation} \label{qubittau}
\tau(\psi) = 4|ad - bc|^2.
\end{equation}
For this simple system, any function of $|\psi\rangle$ that is invariant under local unitary transformations must be a function 
of $\tau(\psi)$.  We can write $\tau(\psi)$ in other ways:
\begin{equation} \label{otherways}
\tau(\psi) = 4\det \rho_{\mathcal A} = 2[1 - \hbox{tr}(\rho_{\mathcal A}^2)] ].
\end{equation}
Here $\rho_{\mathcal A}$ is the reduced density matrix of qubit ${\mathcal A}$, and we could write similar expressions for qubit ${\mathcal B}$.   
The tangle (for a pair of qubits) ranges between the values 0 and 1, with the value 1 corresponding to a maximally entangled state and 0 corresponding
to a product state.  

For a mixed state $\rho$ of two qubits, the tangle is defined as the average tangle of all the pure states in a decomposition of $\rho$,
minimized over all decompositions (this is the convex roof construction) \cite{Osborne}:\footnote{Throughout this paper we use ``min'' rather than ``inf'' in
our expressions for convex roofs.  In every case we consider, the Hilbert space has finite dimension
and the pure-state function to be extended is continuous.  Under these conditions Proposition 3.5 of Ref.~\cite{UhlmannEntropy} guarantees that 
there exists a decomposition achieving the infimum.}
\begin{equation}  \label{mixedtau}
\tau(\rho) = \min \sum_j p_j \tau(\psi_j), \hspace{1cm} \sum_j p_j |\psi_j\rangle\langle \psi_j | = \rho.
\end{equation}
There exists an analytic formula for the tangle of any two-qubit density matrix $\rho$; that is, it is possible to do the 
minimization in Eq.~(\ref{mixedtau}) once and for all   \cite{Hill, Wootters, Osborne}.   The formula is
\begin{equation}   \label{qubittauformula}
\tau(\rho) = [\max\{\lambda_1 - \lambda_2 - \lambda_3 - \lambda_4, 0\}]^2,
\end{equation}
where the $\lambda_j$'s are the square roots of the eigenvalues, in decreasing order, of the matrix
$\rho\tilde{\rho}$.  Here the tilde represents the ``spin-flip'' operation on both qubits:
\begin{equation}
\tilde{\rho} = (Y \otimes Y) \rho^* (Y \otimes Y),
\end{equation}
where the asterisk indicates complex conjugation in the standard basis, and the Pauli matrix
$Y$, written in the same basis, is
\begin{equation}
Y = \mtx{cc}{0 & -i \\ i & 0}.
\end{equation}  

Now consider a system of three qubits, ${\mathcal A}{\mathcal B}{\mathcal C}$, assumed to be in a pure state.  Because ${\mathcal A}$ has a two-dimensional Hilbert space
and the whole system is in a pure state, only two of the four dimensions of the Hilbert space of ${\mathcal B}{\mathcal C}$ will actually be occupied by the
state of ${\mathcal A}{\mathcal B}{\mathcal C}$.  So for the purpose of computing the entanglement between ${\mathcal A}$ and the pair ${\mathcal B}{\mathcal C}$, we may treat ${\mathcal B}{\mathcal C}$ as if it were
a single qubit.  Therefore we can immediately use the definition given in Eqs.~(\ref{qubittau}) and (\ref{otherways}) and write
\begin{equation}
\tau_{{\mathcal A}|{\mathcal B}{\mathcal C}} = 2[1 - \hbox{tr}(\rho_{\mathcal A}^2)].
\end{equation}
(As it happens, the same formula is used for the generalization to higher-dimensional systems \cite{Caves}.)  We can also trace
out qubit ${\mathcal B}$ or ${\mathcal C}$ and compute the tangle between ${\mathcal A}$ and the remaining qubit, thereby defining $\tau_{{\mathcal A}|{\mathcal B}}$ and $\tau_{{\mathcal A}|{\mathcal C}}$.
One can show that the sum of these last two quantities never exceeds $\tau_{{\mathcal A}|{\mathcal B}{\mathcal C}}$ \cite{Coffman}:
\begin{equation}  \label{inequality1}
\tau_{{\mathcal A}|{\mathcal B}{\mathcal C}} \ge \tau_{{\mathcal A}|{\mathcal B}} + \tau_{{\mathcal A}|{\mathcal C}}.
\end{equation}
This equation expresses a version of entanglement monogamy for the case of three qubits.  For example, the left-hand side can never
exceed 1, so there is a trade-off between ${\mathcal A}$'s entanglement with ${\mathcal B}$ and its entanglement with ${\mathcal C}$.  The part of $\tau_{{\mathcal A}|{\mathcal B}{\mathcal C}}$ not 
accounted for in $\tau_{{\mathcal A}|{\mathcal B}}$ and $\tau_{{\mathcal A}|{\mathcal C}}$ is what we define to be the three-tangle 
$\tau_{{\mathcal A}{\mathcal B}{\mathcal C}}$ for a pure state of three qubits:\footnote{This definition appears to be the most widely used, as in 
the review articles \cite{Horodecki, Guhnereview},
%Refs.~\cite{Dur, Verstraete, earlyOsterloh, earlyEltschka, earlyLohmayer, Sharma, Gour1, Hyllus} among others, but in some papers
but in some papers
\cite{lateEltschka1, lateEltschka2, Wu}
the term ``three-tangle'' is used to refer to the {\em square root} of the quantity in Eq.~(\ref{threetangledef}), on the 
basis of arguments given
in Ref.~\cite{Viehmann}.  I prefer not to use the square root in the present
paper because the real-vector-space
analog can be negative.}
\begin{equation}  \label{threetangledef}
\tau_{{\mathcal A}{\mathcal B}{\mathcal C}} = \tau_{{\mathcal A}|{\mathcal B}{\mathcal C}} -  
\tau_{{\mathcal A}|{\mathcal B}} - \tau_{{\mathcal A}|{\mathcal C}}.
\end{equation}
According to Eq.~(\ref{inequality1}), the three-tangle 
$\tau_{{\mathcal A}{\mathcal B}{\mathcal C}}$ is guaranteed to be non-negative.  In fact, it is not hard to see that it can take any value
in the interval $[0,1]$ and no value outside this interval.  (The range of values $0 \le \tau_{ABC} \le 1$ is achieved by states of the form $|\psi\rangle = \cos\theta |000\rangle + \sin\theta |111\rangle$, 
for which $\tau_{\mathcal A \mathcal B \mathcal C} = \tau_{\mathcal A|\mathcal B\mathcal C} = \sin^2(2\theta)$.)

One can work out an explicit formula for the three-tangle in terms of the
components of the state vector \cite{Coffman}.  Let us write this vector as
\begin{equation}
|\psi\rangle = \sum_{ijk} a_{ijk}|ijk\rangle,
\end{equation}
where each index is associated with one of the qubits and takes the values 0 and 1.  Then the three-tangle comes out to be
\begin{equation}  \label{complexthreetangle}
\tau_{{\mathcal A}{\mathcal B}{\mathcal C}} = 4|d_1 - 2d_2 +4d_3|,
\end{equation}
where the $d_j$'s are defined in terms of the components of the state vector as follows:
\begin{equation} \label{ds}
\begin{split}
d_1 &=  a_{000}^2 a_{111}^2 + a_{001}^2 a_{110}^2 + a_{010}^2 a_{101}^2 + a_{100}^2 a_{011}^2  \\
d_2 &=  a_{000}a_{111}a_{011}a_{100} + a_{000}a_{111}a_{101}a_{010} \\
        &\: + a_{000}a_{111}a_{110}a_{001} + a_{011}a_{100}a_{101}a_{010}  \\
        &\; + a_{011}a_{100}a_{110}a_{001} + a_{101}a_{010}a_{110}a_{001} \\
d_3 &= a_{000}a_{110}a_{101}a_{011} + a_{111}a_{001}a_{010}a_{100}
\end{split}
\end{equation}
If we picture the components $a_{ijk}$ lying on the corners of a cube, with each dimension of the cube corresponding
to one of the three indices, then the $d_j$'s consist of all products of four
factors of $a_{ijk}$ whose ``center of mass" is the center of the cube.  Such products fall into three classes, which can also be pictured
geometrically: the terms in $d_1$ correspond to body diagonals, those in $d_2$ to diagonal planes, and those in $d_3$ to 
tetrahedra.  Each of the $d_j$'s is symmetric under interchange of the qubit labels, so the three-tangle itself has this symmetry.  
That is, even though the defining equation (\ref{threetangledef}) assigns qubit $A$ a special role, the three-tangle does not 
depend on this choice, as we noted in the Introduction.  The special combination of $d_j$'s that appears in our expression for
$\tau_{{\mathcal A}{\mathcal B}{\mathcal C}}$ has a fairly long history in mathematics: without the absolute value sign, the combination $d_1 - 2d_2 + 4d_3$ is the Cayley hyperdeterminant (introduced in 1845) of the three-index array $a_{ijk}$ \cite{Cayley, Miyake1, Miyake2}.

\section{The analog of the three-tangle for rebits}

We now want to follow exactly the same line of argument as above, but in the context of the real-vector-space variant of quantum theory.
That is, we will take all state vectors to be real.  This is the only change.  For clarity we will use the symbol $\sigma$ instead of $\tau$
when we are referring to tangles and three-tangles in the real-vector-space theory, but the definitions are essentially the same.  The definition of the tangle for a pure state is 
still given by Eq.~(\ref{qubittau})---the forms appearing in Eq.~(\ref{otherways}) are still valid---and the tangle for a mixed state is still given by Eq.~(\ref{mixedtau}).  (Imagine a $\sigma$ in place of
each $\tau$ in these equations.)  However, in doing the minimization 
called for in Eq.~(\ref{mixedtau}), we are now allowed only those decompositions consisting of pure states whose state vectors are real.  This 
restriction makes an important difference, and the tangle of a mixed state of two rebits turns out not to be given by Eq.~(\ref{qubittauformula}).
Fortunately, for the case of two rebits it is again possible to find an analytic expression for the tangle of a mixed state.  The essential
argument is given in Ref.~\cite{Rungta}.  The authors of that paper actually found the entanglement of formation of a pair of rebits,
but the same argument gives us the tangle.\footnote{In Ref.~\cite{Rungta}, the only property of 
entanglement of formation needed in the proof was that for a pure state,
it is a convex and monotonic function of the concurrence.  But the tangle, too, is a convex, monotonic function of concurrence for a pure state:
it is simply the square of the concurrence.  So the same argument applies.}  It is
\begin{equation}  \label{rebittauformula}
\sigma(\rho) = \big(\hbox{tr}[(Y \otimes Y)\rho]\big)^2.
\end{equation}
Notice that this quantity is much easier to compute than the two-qubit tangle given by Eq.~(\ref{qubittauformula}).  Rather than having 
to find the eigenvalues of a matrix, one needs only to compute a trace.  

We can illustrate the difference between Eq.~(\ref{qubittauformula}) and Eq.~(\ref{rebittauformula}) for a specific state.  
Consider the real density matrix
\begin{equation}  \label{examplestate}
\rho = (1/4)[I \otimes I + Y \otimes Y],
\end{equation}
where $I$ is the $2\times 2$ identity matrix.  One can check that, when we regard this state as a state of two qubits, Eq.~(\ref{qubittauformula})
tells us that the tangle is zero, whereas when we regard it as a state of two rebits, Eq.~(\ref{rebittauformula}) tells us that the tangle is 1.  
To see where the difference comes from, note that, when regarded as a state of two qubits, 
$\rho$ can be written as an equal mixture of the two pure states
\begin{equation}  \label{twoproductstates}
[(I + Y)/2] \otimes [(I + Y)/2] \hspace{2mm}\hbox{and}\hspace{2mm}
[(I - Y)/2] \otimes [(I - Y)/2] .
\end{equation}
Each of these states is a product state, so the state $\rho$ is unentangled, being a mixture of two product states.  However,
these product states are complex and are therefore not allowed in the real-vector-space theory.  Indeed, one can show that in the real-vector-space
theory, {\em every} decomposition of $\rho$ into pure states consists of maximally entangled states.  
Hence there is no minimizing to be done: the average tangle appearing in Eq.~(\ref{mixedtau}) has the value 1 for every decomposition.  

We now define the three-tangle $\sigma_{ABC}$ for a pure state of three rebits, using the same defining relation we used for qubits:
\begin{equation} \label{threetangleagain}
\sigma_{ABC} = \sigma_{A|BC} - \sigma_{A|B} - \sigma_{A|C}.
\end{equation}
%Note that $\sigma_{ABC}$, and indeed each term on the right-hand side of Eq.~(\ref{threetangleagain}), is invariant under a local rotation
%of any of the rebits.  
But now the mixed-state tangles $\sigma_{A|B}$ and $\sigma_{A|C}$ are to be given by Eq.~(\ref{rebittauformula}) and not by
Eq.~(\ref{qubittauformula}), and this difference will make a difference in our formula for $\sigma_{ABC}$.
Let us write the pure state as 
\begin{equation}
|\phi\rangle = \sum_{ijk} a_{ijk}|ijk\rangle,
\end{equation}
where the binary indices $i,j,k$ correspond to rebits $A$, $B$, and $C$, respectively.  We want to write each of the terms on the 
right-hand side of Eq.~(\ref{threetangleagain}) in terms of the real numbers $a_{ijk}$.  

To get $\sigma_{A|BC}$, we write down the reduced density matrix $\rho_A$:
\begin{equation}
(\rho_A)_{ii'} = \sum_{jk} a_{ijk}a_{i'jk}.
\end{equation} 
It is probably easiest to compute $\sigma_{A|BC}$ by writing it as $\sigma_{A|BC} = 4\det\rho_A$, which gives us
\begin{equation}  \label{ABC}
\begin{split}
\sigma_{A|BC} = 4 &\left[ (a_{000}^2 + a_{001}^2 + a_{010}^2 + a_{011}^2)(a_{100}^2 + a_{101}^2 + a_{110}^2 + a_{111}^2) \right. \\
& \left. - (a_{000}a_{100} + a_{001}a_{101} + a_{010}a_{110} + a_{011}a_{111})^2  \right]
\end{split}
\end{equation}
%A bit of algebra then gives us
%\begin{equation} \label{ABC}
%\begin{split}
%\sigma_{A|BC} = 4[&(a_{000}a_{101} - a_{100}a_{001})^2 + (a_{000}a_{110}-a_{100}a_{010})^2 + (a_{000}a_{111} - a_{100}a_{011})^2 \\
%          + & \; (a_{001}a_{110}-a_{101}a_{010})^2 + (a_{001}a_{111} - a_{101}a_{011})^2 + (a_{010}a_{111} - a_{110}a_{011})^2].
%\end{split}
%\end{equation}
To get $\sigma_{A|B}$, we need $\rho_{AB}$, whose components are
\begin{equation}
(\rho_{AB})_{ij,i'j'} = \sum_k a_{ijk}a_{i'j'k}.
\end{equation}
Then Eq.~(\ref{rebittauformula}) gives us
%In fact we need only two components of $\rho_{AB}$, because from Eq.~(\ref{rebittauformula}) we have
%\begin{equation}
%\sigma_{A|B} = \big(\hbox{tr}[(Y \otimes Y)\rho_{AB}]\big)^2 = 4[(\rho_{AB})_{01,10} - (\rho_{AB})_{00,11}]^2.
%\end{equation}
%Here we have used the fact that $\rho_{AB}$ is symmetric, so that $(\rho_{AB})_{10,01} = (\rho_{AB})_{01,10}$
%and $(\rho_{AB})_{11,00} = (\rho_{AB})_{00,11}$.  Writing out the two relevant components, we get
\begin{equation} \label{AB}
\sigma_{A|B} = 4[a_{010}a_{100} + a_{011}a_{101} - a_{000}a_{110} - a_{001}a_{111}]^2.
\end{equation}
By interchanging the last two indices in each factor, we immediately also get
\begin{equation} \label{AC}
\sigma_{A|C} = 4[a_{001}a_{100} + a_{011}a_{110} - a_{000}a_{101} - a_{010}a_{111}]^2.
\end{equation}
It is now not hard to combine Eqs.~(\ref{ABC}), (\ref{AB}), and (\ref{AC}) to find the rebit three-tangle, which comes out to be
\begin{equation} \label{realthreetangle}
\sigma_{ABC} = \sigma_{A|BC} - \sigma_{A|B} - \sigma_{A|C} = 4(d_1 - 2d_2 +4d_3),
\end{equation}
where the $d_j$'s are given as before by Eq.~(\ref{ds}).

Thus the rebit three-tangle differs from the standard three-tangle only by the absence of the absolute value sign.  That is,
the rebit three-tangle is simply four times the Cayley hyperdeterminant of $a_{ijk}$.  

Should we be surprised
that the formula for $\sigma_{ABC}$ is so similar to the one for $\tau_{\mathcal A\mathcal B\mathcal C}$? 
On the one hand, the hyperdeterminant was certainly
a plausible candidate for the rebit three-tangle, since it is invariant under local rotations.  (Note that all the quantities 
appearing in Eq.~(\ref{threetangleagain})---that is,  
$\sigma_{A|BC}$, $\sigma_{A|B}$, $\sigma_{A|C}$, and therefore $\sigma_{ABC}$---are invariant under local
rotations.)  On the other hand, I do not think it was obvious
at the outset that $\sigma_{ABC}$ would be invariant under permutations of the rebits, as Eq.~(\ref{realthreetangle}) shows 
that it is.   
We could not have deduced this fact from the similar invariance of the ordinary three-tangle, since $\tau_{\mathcal A\mathcal B\mathcal C}$
was derived from Eq.~(\ref{qubittauformula}), which is notably different from the analogous rebit formula
in Eq.~(\ref{rebittauformula}).
At any rate, the invariance under permutations shows us that $\sigma_{ABC}$ is
characteristic of the triple of rebits as a whole, with
no bias toward any one rebit.  We can also conclude from Eq.~(\ref{realthreetangle}) that $\sigma_{ABC}$ is confined to the interval $[-1,1]$, since the ordinary
three-tangle is confined to the interval $[0,1]$.

As we noted in the Introduction, it happens that because of the absence of the absolute value sign, the rebit three-tangle can 
indeed be negative.  Consider, for example, the following pure state of three rebits:
\begin{equation}  \label{phi}
|\phi\rangle = (1/2)[|000\rangle - |011\rangle - |101\rangle - |110\rangle ].
\end{equation}
In this state, rebit $A$ is maximally entangled with the pair $BC$, so $\sigma_{A|BC}$ is equal to 1.  When we trace out rebit $C$, we find that the remaining
two rebits have the density matrix $\rho_{AB} = (1/4)[I \otimes I + Y \otimes Y]$, which we saw earlier has a tangle of 1.  
So $\sigma_{A|B} = 1$.  The pair $AC$ is described by the same density matrix, so $\sigma_{A|C}$ is also equal to $1$.  Therefore, by the definition (\ref{threetangleagain}), the rebit three-tangle 
must be negative:
\begin{equation}
\sigma_{ABC} = \sigma_{A|BC} - \sigma_{A|B} - \sigma_{A|C} = 1 - 1 - 1 = -1.
\end{equation}
And indeed, the formula given in Eq.~(\ref{realthreetangle}) produces the value $-1$ for the state $|\phi\rangle$: the quantities $d_1$ and $d_2$ are zero while
$d_3$ has the value $-1/16$.  This state $|\phi\rangle$ serves as a good example illustrating the lack of entanglement monogamy in the
real-vector-space theory \cite{Wootters2}.  Along with $\sigma_{A|B}$ and $\sigma_{A|C}$, the tangle $\sigma_{B|C}$ is also equal to 1, so that each pair of rebits 
is in a maximally entangled state.  I should emphasize, however, that this lack of monogamy applies to entanglement as measured by 
the tangle, which is based on the convex-roof construction.  There may well be other notions of 
entanglement that are perfectly monogamous even in the real-vector-space theory.  

The possibility of negative values also makes it clear that the rebit three-tangle $\sigma_{ABC}$ is not a monotone under local operations and 
classical communication (LOCC).  Imagine starting with the state $|\phi\rangle$ of Eq.~(\ref{phi}) and then doing a projective
measurement on each rebit in the basis $\{|0\rangle, |1\rangle \}$.
In this case the initial state has $\sigma_{ABC} = -1$, but each of the four possible final states is a pure product state with $\sigma_{ABC} = 0$.
Thus one can increase the rebit three-tangle locally.   (In fact, from the state $|\phi\rangle$, a projective measurement on just one of the three rebits is sufficient to raise the average of $\sigma_{ABC}$ to the value zero.)  
On the other hand, the {\em absolute value} of $\sigma_{ABC}$ is an LOCC monotone in the real-amplitude
theory, since it is equal to the standard three-tangle, which is an LOCC monotone in the complex theory \cite{DurVidalCirac}, and since every local operation on rebits can also be regarded as
a local operation on qubits.  
%It is also conceivable that one could develop a resource theory in which both $\sigma_{ABC}$ and local measurements are counted as resources---one can increase
%$\sigma_{ABC}$ locally only by paying for the increase by performing a measurement.  However, we do not pursue that idea in this paper.  

So far we have defined the rebit three-tangle $\sigma_{ABC}$ only for pure states.  Later we will want to apply the concept to mixed states as well, so we need 
to extend our definition.  For the standard three-tangle, the extension is given by the convex roof construction \cite{earlyLohmayer}, and we use the same construction here.  That is,
for a mixed state $\rho$ of three rebits, we define the rebit three-tangle to be
\begin{equation}  \label{defmixedthreetangle}
\sigma_{ABC}(\rho) = \min \sum_j p_j \sigma_{ABC}(\phi_j), \hspace{3mm} \sum_j p_j |\phi_j\rangle\langle \phi_j | = \rho.
\end{equation}
This definition deserves some comment.  One of the motivations often given for the convex roof construction is that it preserves 
LOCC monotonicity \cite{PlenioVirmani}.  This motivation does not apply to $\sigma_{ABC}$ since it is not an LOCC monotone.  
However, the convex roof construction has a number of other special features that make it suitable as a way of extending to mixed states
a continuous function defined on pure states \cite{UhlmannEntropy, Hellmund}.
In particular, for any given pure-state function, the function defined by the convex roof construction is distinguished as the largest of all the convex extensions of the given function \cite{UhlmannEntropy, Hellmund, Rockafellar}.  In this sense
it captures as much of the pure-state information as possible without ever assigning to a mixture of states a value greater than the average
value on the states being mixed.  Even though $\sigma_{ABC}$ is not an LOCC monotone, one can imagine it figuring 
into a different kind of resource theory---e.g., a theory in which a cost is associated with local measurements---in which case one would presumably still want
it to be non-increasing on average under the operation of mixing.  The definition given in Eq.~(\ref{defmixedthreetangle}) guarantees at least this kind
of monotonicity.

\section{Writing a two-qubit state as a three-rebit state}

Given any pure or mixed state $\rho_{\mathcal S}$ of a quantum system ${\mathcal S}$ with Hilbert-space dimension $d$, one can always re-express the state in terms of the real-vector-space
variant of quantum theory, in the following way \cite{Stueckelberg, Myrheim, Aleksandrova, Mosca}.  First we replace $\mathcal S$ with 
a pair of objects: (i) a real-vector-space quantum object $S$ also having a $d$-dimensional Hilbert space (but over the reals), and (ii)
a rebit $U$.  We call $U$ the universal rebit, or ubit, because a single such rebit is all that is needed, no matter how many component 
systems $\mathcal S$ might contain \cite{Aleksandrova}.  (We do not use a separate $U$ for each elementary subsystem.  For a model with separate ancillas see Ref.~\cite{Mosca}.)  The real version of $\rho_{\mathcal S}$ can then be written as
\begin{equation}  \label{rhoUS}
\rho_{US} = (1/2)[I_U \otimes  \left(\hbox{Re}\, \rho_{\mathcal S} \right)_S +J_U \otimes \left( \hbox{Im}\, \rho_{\mathcal S}\right)_S],
\end{equation}
where $J$ is the $2 \times 2$ matrix
\begin{equation}
J = \mtx{cc}{0   &  -1  \\ 1  &  0}
\end{equation}
and the subscripts $U$ and $S$ refer to the two subsystems.  Note that $\rho_{US}$ is a symmetric real matrix with unit trace.  It is in fact
a legitimate density matrix in the real-vector-space theory.  By similarly mapping all measurement and transformation operators into 
operators on the $US$ system, one can rewrite all of quantum theory in real-vector-space terms \cite{Stueckelberg, Myrheim, Aleksandrova, Mosca}.  For example, if a measurement outcome
is associated with a projection operator $\Pi$ in the complex theory, and if the system being measured is in the state $\rho_{\mathcal S}$, 
then the probability of that outcome, $\hbox{tr}\left(\Pi\rho_{\mathcal S}\right)$, can alternatively be written as 
\begin{equation}
\hbox{tr}\left(\Pi\rho_{\mathcal S}\right) = \hbox{tr}\left(P \rho_{US}\right),
\end{equation}
where 
$P$ is the real projection operator $I_U \otimes \left(\hbox{Re}\,\Pi\right)_S + J_U \otimes \left( \hbox{Im}\,\Pi\right)_S$
and $\rho_{US}$ is given by Eq.~(\ref{rhoUS}).
But here our focus is simply on the
states.  

When $\rho_{\mathcal S}$ is a pure state $\rho_{\mathcal S} = |\psi\rangle \langle \psi |$, then the corresponding $\rho_{US}$ is a rank-two
density matrix whose nonzero eigenvalues are $1/2$ and $1/2$, as we now show  
by writing down an explicit decomposition of $\rho_{US}$.  Let $|\psi\rangle = |a\rangle + i|b\rangle$, where
$|a\rangle$ and $|b\rangle$ are real.  (The vectors $|a\rangle$ and $|b\rangle$ are not individually normalized but satisfy
$\langle a | a\rangle + \langle b |b\rangle = 1$.)  Then $\rho_{US}$ is an equal mixture of the following two 
orthogonal pure states (they are orthogonal regardless of the relation between $|a\rangle$ and $|b\rangle$):
\begin{equation}  \label{phi12}
\begin{split}
&|\xi_1\rangle = |0\rangle_U \otimes |a\rangle_S + |1\rangle_U \otimes |b\rangle_S \\
&|\xi_2\rangle = J_U|\xi_1\rangle =  -|0\rangle_U \otimes |b\rangle_S + |1\rangle_U \otimes |a\rangle_S,
\end{split}
\end{equation}
where we are using $J_U$ as shorthand for $J_U \otimes I_S$.
Indeed, by writing out $(1/2)(|\xi_1\rangle\langle \xi_1| + |\xi_2\rangle\langle \xi_2|)$ with the expressions given in 
Eq.~(\ref{phi12}), one directly obtains the form given in Eq.~(\ref{rhoUS}).

We now specialize to the case in which the system $\mathcal S$ is a pair of qubits $\mathcal A \mathcal B$.  Let these qubits be 
in a pure state $|\psi\rangle = |a\rangle + i|b\rangle$, where again $|a\rangle$ and $|b\rangle$ are real vectors.  According to the
above correspondence, the associated
state of the three-rebit system $UAB$ is $\rho_{UAB} = (1/2)(|\xi_1\rangle\langle \xi_1| + |\xi_2\rangle\langle \xi_2|)$, where
\begin{equation}  \label{phi12UAB}
\begin{split}
&|\xi_1\rangle = |0\rangle_U \otimes |a\rangle_{AB} + |1\rangle_U \otimes |b\rangle_{AB}  \\
&|\xi_2\rangle = J_U|\xi_1\rangle =  -|0\rangle_U \otimes |b\rangle_{AB} + |1\rangle_U \otimes |a\rangle_{AB}.
\end{split}
\end{equation}
In terms of $|a\rangle$ and $|b\rangle$, the density matrix $\rho_{UAB}$ can be written as
\begin{equation}
\begin{split}
\rho_{UAB} = (1/2)\big[ &I_U \otimes \left( |a\rangle\langle a| + |b\rangle\langle b| \right)_{AB} \\
 + &J_U \otimes \left( |b\rangle\langle a| - |a\rangle\langle b |\right)_{AB}\big].
\end{split}
\end{equation}
In the following section, we compute the tangle $\sigma_{A|B}$ and the three-tangle $\sigma_{UAB}$ for this three-rebit state and relate these 
quantities 
to the tangle between the two {\em qubits} $\mathcal A$ and $\mathcal B$.

\section{Relating the $\mathcal A\mathcal B$ tangle to the $AB$ tangle}

Given the above correspondence between a two-qubit pure state and a three-rebit rank-2 mixed state, we show in this
section that the two-qubit tangle $\tau_{\mathcal A|\mathcal B}$ and the two-rebit tangle $\sigma_{A|B}$ differ from each other
precisely by the rebit three-tangle.  Specifically,
\begin{equation}  \label{2to3}
\tau_{\mathcal A|\mathcal B} = \sigma_{A|B} + \sigma_{UAB}.
\end{equation}
To prove this, let us write down expressions for the three terms appearing in Eq.~(\ref{2to3}), in the order in which they appear.  

Again writing the state of the two qubits as $|\psi\rangle = |a\rangle + i|b\rangle$, with $|a\rangle$ and $|b\rangle$ real,
we can get the two-qubit tangle $\tau_{\mathcal A|\mathcal B}$ from Eq.~(\ref{qubittau}):
\begin{equation}
\tau_{\mathcal A|\mathcal B} = 4\big|(a_{00} + ib_{00})(a_{11} + ib_{11}) - (a_{01}+ ib_{01})(a_{10} + ib_{10})\big|^2,
\end{equation}
where the components $a_{jk}$ are defined by $|a\rangle = \sum_{jk}a_{jk}|jk\rangle$, and similarly for $b_{jk}$.  Separately squaring the
real and imaginary parts of the expression inside $|\cdots|$, we obtain
\begin{equation}  \label{twoqubittau}
\tau_{\mathcal A|\mathcal B} = \big[ \langle a| J \otimes J |a\rangle - \langle b | J \otimes J |b\rangle\big]^2
+ 4 \langle a | J \otimes J | b\rangle^2.
\end{equation}
To get the two-rebit tangle $\sigma_{A|B}$, we trace $U$ out of $\rho_{UAB}$ to get $\rho_{AB}$, and then use 
the formula (\ref{rebittauformula}).  Noting that the operator $Y \otimes Y$ appearing in Eq.~(\ref{rebittauformula})
can also be written as $-J\otimes J$, we find that
\begin{equation}  \label{tworebittau}
\sigma_{A|B} = \big[ \langle a|J \otimes J |a\rangle + \langle b |J \otimes J |b\rangle \big]^2.
\end{equation}

Finally we need the three-tangle $\sigma_{UAB}$.  The state $\rho_{UAB}$ is mixed,
so we use the definition of the rebit three-tangle given in Eq.~(\ref{defmixedthreetangle}):
\begin{equation} 
\sigma_{UAB}(\rho) = \min \sum_j p_j \sigma_{UAB}(\phi_j), \hspace{3mm} \sum_j p_j |\phi_j\rangle\langle \phi_j | = \rho.
\end{equation}
In fact, however, all decompositions of our rank-2 state $\rho_{UAB}$ yield the same average value of $\sigma_{UAB}$.  To see this, note that
any pure state $|\phi\rangle$ in the support of $\rho_{UAB}$ must be a linear
combination of
$|\xi_1\rangle$ and $|\xi_2\rangle$ as given in Eq.~(\ref{phi12UAB}):
\begin{equation}  \label{onlyarotation}
\begin{split}
|\phi\rangle &= \cos\theta|\xi_1\rangle + \sin\theta|\xi_2\rangle \\
                     &= \left[(\cos\theta) I_U + (\sin\theta) J_U \right] |\xi_1\rangle \\
                     &= R_U |\xi_1\rangle.
\end{split}
\end{equation}
Here $R$ is the rotation matrix
\begin{equation}
R = \mtx{cc}{\cos\theta & -\sin\theta \\ \sin\theta & \cos\theta}.
\end{equation}
That is, any such $|\phi\rangle$ is obtained from $|\xi_1\rangle$ simply by rotating the ubit.  But this rotation is a local transformation and therefore does not
affect $\sigma_{UAB}$.  
So 
again there is no minimization to be done, and $\sigma_{UAB}(\rho_{UAB})$ will simply be equal to the three-tangle of every pure state in any decomposition of $\rho_{UAB}$.   

Let us compute $\sigma_{UAB}$, then, for the specific
pure state $|\xi_1\rangle$, starting with  
the definition of $\sigma_{UAB}$:
\begin{equation}  \label{UAB}
\sigma_{UAB} = \sigma_{U|AB} - \sigma_{U|A} - \sigma_{U|B}.
\end{equation}
We begin by computing $\sigma_{U|AB}$.   Tracing $|\xi_1\rangle\langle \xi_1 |$ over $A$ and $B$, we find that
\begin{equation}
\rho_U = \mtx{cc}{\langle a | a \rangle & \langle a | b\rangle \\ \langle a | b \rangle & \langle b | b \rangle},
\end{equation}
and therefore
\begin{equation}
\sigma_{U|AB}(\xi_1) = 4 \det \rho_U = 4\left(\langle a|a\rangle \langle b|b\rangle - \langle a|b\rangle^2\right).
\end{equation}
We obtain $\sigma_{U|A}$ and $\sigma_{U|B}$ for the state $|\xi_1\rangle$ from the formula (\ref{rebittauformula}), which
gives us
\begin{equation}
\begin{split}
&\sigma_{U|A}(\xi_1) =4 \langle a|J \otimes I|b\rangle^2   \\
 \hbox{and} \hspace{4mm}
&\sigma_{U|B}(\xi_1) =4 \langle a|I \otimes J|b\rangle^2.
\end{split}
\end{equation}
Thus the three-tangle for this pure state is
\begin{equation}  \label{UAB3}
\begin{split}
\sigma_{UAB}(\xi_1) &= 4\left[\left(\langle a|a\rangle \langle b|b\rangle  \right.
 - \langle a|b\rangle^2\right) \\
 &\;\;\;\;\;\; \left.-\langle a|J \otimes I|b\rangle^2 - \langle a|I \otimes J|b\rangle^2\right].
\end{split}
\end{equation}
According to what we have said above, this quantity is also the three-tangle for the rank-2 state $\rho_{UAB}$.

We now put the pieces together.  According to Eqs.~(\ref{twoqubittau}) and (\ref{tworebittau}), the difference between 
$\tau_{\mathcal A|\mathcal B}$ and $\sigma_{A|B}$ is
\begin{equation} \label{xxx}
\begin{split}
\tau_{\mathcal A|\mathcal B} - \sigma_{A|B} &= 
 \big[ \langle a| J \otimes J |a\rangle - \langle b | J \otimes J |b\rangle\big]^2 \\
&+ 4 \langle a | J \otimes J | b\rangle^2
-\big[ \langle a|J \otimes J |a\rangle + \langle b |J \otimes J |b\rangle \big]^2 \\
&= 4\big[ \langle a | J \otimes J | b\rangle^2 - \langle a | J \otimes J | a\rangle \langle b | J \otimes J | b\rangle \big].
\end{split}
\end{equation}
Our claim is that this is the same as $\sigma_{UAB}$, given in Eq.~(\ref{UAB3}).
The equivalence between Eqs.~(\ref{UAB3}) and (\ref{xxx}) follows from the (non-obvious) identity
\begin{equation} \label{identity}
\begin{split}
&\hspace{5mm}\langle a|a\rangle \langle b|b\rangle + \langle a |J \otimes J|a\rangle \langle b|J \otimes J |b\rangle \\
&= \langle a|b\rangle^2 + \langle a| I \otimes J |b\rangle^2 + \langle a| J \otimes I |b\rangle^2 + \langle a| J \otimes J |b\rangle^2.
\end{split}
\end{equation}
One can prove this identity by expanding the operator $|b\rangle\langle a|$ in the orthonormal basis of operators
$(1/2)\{I, J, X, Z\} \otimes \{I, J, X, Z\}$.  Here $X$ and $Z$ are Pauli matrices, and the orthonormality is
with respect to the 
Hilbert-Schmidt inner product.  Let us call these basis operators $B_j$, $j = 1,\ldots, 16$, and let the $j$th component
of $|b\rangle\langle a|$ be $\alpha_j$.  That is,
\begin{equation}
\begin{split}
&|b\rangle\langle a| = \sum_j \alpha_j B_j, \\ 
\hbox{where} \hspace{4mm} &\alpha_j = \hbox{tr}\left( B_j^T|b\rangle\langle a | \right)
= \langle a | B_j^T |b\rangle.
\end{split}
\end{equation}
Then
\begin{equation}
\langle a|a\rangle\langle b|b\rangle = \hbox{tr}\left( |b\rangle\langle a|a\rangle \langle b| \right)
= \sum_j \alpha_j^2
\end{equation}
and similarly
\begin{equation}
\begin{split}
&\hspace{4mm}\langle a| J \otimes J |a \rangle\langle b| J \otimes J |b\rangle \\
&= \hbox{tr} \left( |b\rangle \langle a| J\otimes J |a\rangle\langle b | J \otimes J \right) \\
&= \sum_j \alpha_j^2 s_j,
\end{split}
\end{equation}
where $s_j$ equals $+1$ if $B_j$ commutes with $J \otimes J$ and $-1$ if it anticommutes.  Thus
\begin{equation}  \label{dddd}
\langle a|a\rangle\langle b|b\rangle + \langle a| J \otimes J |a \rangle\langle b| J \otimes J |b\rangle
= 2 \sum_{j \in S_8} \alpha_j^2,
\end{equation}
where $S_8$ is the set of eight values of $j$ for which $B_j$ commutes with $J \otimes J$.  Finally, we use the 
fact that $I \otimes J$ is antisymmetric, which implies that $\langle a|I\otimes J |a\rangle = 0$ and therefore
\begin{equation}
\begin{split}
0&=\langle a| I \otimes J |a\rangle\langle b| I \otimes J |b\rangle \\
&= \hbox{tr} \left( |b\rangle \langle a| I\otimes J |a\rangle\langle b | I \otimes J \right)
= -\sum_j \alpha_j^2 t_j,
\end{split}
\end{equation}
where $t_j$ equals $+1$ when $B_j$ commutes with $I\otimes J$ and $-1$ otherwise.  This last relation, together with 
a similar relation obtained from the antisymmetry of $J \otimes I$, allows us to rewrite the 
sum in Eq.~(\ref{dddd}) using just the components associated with $B_j$'s that commute with both $J \otimes J$ and $I \otimes J$:
\begin{equation}
\langle a|a\rangle\langle b|b\rangle + \langle a| J \otimes J |a \rangle\langle b| J \otimes J |b\rangle
= 4 \sum_{j \in S_4} \alpha_j^2,
\end{equation}
where $S_4$ is the set of four values of $j$ corresponding to the basis operators $(1/2)\{I, J\} \otimes \{I, J\}$.  This sum gives
us the four terms on the right-hand side of Eq.~(\ref{identity}) and proves the identity.

We have thus shown that 
$\tau_{\mathcal A|\mathcal B} = \sigma_{A|B} + \sigma_{UAB}$, which is the main result of this section.  
This equation gives us an interpretation of the rebit three-tangle when it is applied to a two-qubit pure state 
re-written in the ubit model: it expresses the difference between the original two-qubit entanglement 
and the corresponding two-rebit entanglement.  Note that there is no {\em a priori} ordering of the sizes of these 
two entanglements, since $\sigma_{UAB}$ can be either positive or negative.  

I hasten to add that the two-rebit entanglement $\sigma_{A|B}$ does not have the physical meaning one might expect. 
The role of the ubit is to bring the complex structure into what is otherwise a real vector space.  So in the representation of
a two-qubit state as a three-rebit state, the rebits $A$ and $B$ can be entangled merely by virtue of a separation between
the real and imaginary parts of what would have been a complex vector.  Consider, for example, the two-qubit product state
\begin{equation}
\begin{split}
|\psi\rangle &= (1/2)(|0\rangle + i|1\rangle)\otimes (|0\rangle +i |1\rangle) \\
&=(1/2)(|00\rangle + i|01\rangle + i|10\rangle - |11\rangle).
\end{split}
\end{equation}
(This is the first of the two states in Eq.~(\ref{twoproductstates}).)
In the three-rebit version of this state, the reduced density matrix $\rho_{AB}$ is an equal mixture of 
the states $(1/\sqrt{2})(|00\rangle - |11\rangle)$ and $(1/\sqrt{2})(|01\rangle + |10\rangle)$, which is fully entangled in the
real-vector-space world, though the two qubits are clearly unentangled.  In this case, the rebit three-tangle
$\sigma_{UAB}$ makes up for this real-number-induced entanglement by being negative: for this state $|\psi\rangle$,
Eq.~(\ref{2to3}) reads $0 = 1 + (-1)$.    

In the next two paragraphs, 
we show that we can write the relation $\tau_{\mathcal A|\mathcal B} = \sigma_{A|B} + \sigma_{UAB}$ in a couple of other forms that have a certain intuitive appeal.  

Let us begin by recalling that for a {\em pure} state of three rebits $UAB$, the rebit three-tangle 
can be written in different ways, depending on which rebit one takes as the ``hinge'':
\begin{equation}  \label{threeequations}
\begin{split}
\sigma_{UAB} &= \sigma_{U|AB} - \sigma_{U|A} - \sigma_{U|B} \\
                     &= \sigma_{A|UB} - \sigma_{A|U} - \sigma_{A|B} \\
                     &= \sigma_{B|UA} - \sigma_{B|U} - \sigma_{B|A} 
\end{split}
\end{equation}
It is not hard to see that these equations are also true for the rank-2 three-rebit density matrix $\rho_{UAB}$ derived from a two-qubit pure state. First note that
each term in these equations is defined by minimizing an average value over all decompositions of $\rho_{UAB}$.  But in each case,
the quantity to be averaged takes the same value on {\em every} pure state in the support of $\rho_{UAB}$, because these pure
states are all related to each other by a simple rotation of the ubit, as we saw in Eq.~(\ref{onlyarotation}).  Since the equations (\ref{threeequations}) are true for pure states of $UAB$, they are also true for these particular mixed states.  

We can now combine the last two of the equations in (\ref{threeequations}) with the main result of this section,
$\tau_{\mathcal A|\mathcal B} = \sigma_{A|B} + \sigma_{UAB}$, to arrive at the following relations: for any pure state
of two qubits,
\begin{equation}
\begin{split}
&\tau_{\mathcal A|\mathcal B} = \sigma_{A|UB} - \sigma_{A|U} \\
 \hbox{and} \hspace{4mm}
&\tau_{\mathcal A|\mathcal B} = \sigma_{B|UA} - \sigma_{B|U}.
\end{split}
\end{equation}
Thus to compute the standard tangle between qubits $\mathcal A$ and $\mathcal B$ in a pure state, one can compute the real-amplitude tangle
between rebit $A$ and the pair $UB$, and then subtract off the tangle between $A$ and $U$.  For the example given above,
that is, $|\psi\rangle = (1/2)(|0\rangle + i|1\rangle)\otimes (|0\rangle + i |1\rangle)$, one finds that rebit $A$ is fully entangled with the 
pair $UB$, but it is also fully entangled with $U$ itself.  So the difference, which gives the actual tangle between the qubits
$\mathcal A$ and $\mathcal B$, is zero, as it should be.

\section{Mixed states of two qubits}

We have seen that when a pure state of two qubits is rewritten as a state of three rebits, the original two-qubit tangle can be 
written as
\begin{equation} \label{aaa}
 \tau_{\mathcal A|\mathcal B} = \sigma_{A|B} + \sigma_{UAB}
 \end{equation} 
 or as 
 \begin{equation} \label{bbb}
 \tau_{\mathcal A|\mathcal B} = \sigma_{A|UB} - \sigma_{A|U}.
 \end{equation}
One might wonder whether either of these equations remains true for all {\em mixed} states of two qubits.  The answer in both cases
is no.  

A counterexample to Eq.~(\ref{aaa}) is the completely mixed state $\rho_{\mathcal A \mathcal B} = (1/4)I_{\mathcal A} \otimes I_{\mathcal B}$.  The state is
unentangled, so $\tau_{\mathcal A|\mathcal B}$ is zero.  The equivalent three-rebit state is $\rho_{UAB} = (1/8)I_U \otimes I_A \otimes I_B$,
for which $\sigma_{A|B}$ is also zero.  But perhaps surprisingly, the rebit three-tangle $\sigma_{UAB}$ of this state is $-1$.  
This is because we can decompose $\rho_{UAB}$ into the eight possible ``tetrahedral'' states of the form
\begin{equation}
\begin{split}
&a|000\rangle + b|011\rangle + c|101\rangle + d|110\rangle \\
\hbox{or} \hspace{4mm}
&a|111\rangle + b|100\rangle + c|010\rangle + d|001\rangle,
\end{split}
\end{equation}
where one of the four coefficients $a, b, c, d$ has the value $-1/2$ and the others have the value $+1/2$.  
Each of these eight states has $\sigma_{UAB}$ equal to $-1$ (the smallest value possible).  Since the mixed-state three-tangle
is defined by minimizing over all decompositions, this is also the value of $\sigma_{UAB}$ for the completely mixed state.  

A counterexample to Eq.~(\ref{bbb}) is the mixed state given in
Eq.~(\ref{examplestate}), which again is
\begin{equation}
\rho_{\mathcal A \mathcal B} = (1/4)[I \otimes I + Y \otimes Y].
\end{equation}
We have seen earlier that this is an unentangled state of two qubits, so $\tau_{\mathcal A|\mathcal B} = 0$.  One can also check that
for the three-rebit version of the state,
$\sigma_{A|U} = 0$ and $\sigma_{A|UB} = 1$, 
so that Eq.~(\ref{bbb}) is false.  That $\sigma_{A|UB}$ has the value $1$ can be shown by noting that every pure state in the support of 
$\rho_{UAB}$ has maximal entanglement between the rebit $A$ and the pair $UB$.  So no matter which decomposition one chooses, 
the average value of $\sigma_{A|UB}$ is 1.

Even though Eqs.~(\ref{aaa}) and (\ref{bbb}) do not hold in general, there is a way of extending the results of the preceding section to mixed states.  The following statements, making use
of the convex-roof construction, are both true for a general two-qubit state $\rho_{\mathcal A\mathcal B}$ and the corresponding
three-rebit state $\rho_{UAB}$ given by Eq.~(\ref{rhoUS}) (which now may be of full rank).
\begin{equation} \label{convexroof1}
\begin{split}
\tau_{\mathcal A|\mathcal B} = \min &\sum_j p_j (\sigma_{A|B}(\phi_j) + \sigma_{UAB}(\phi_j)), \\
 &\sum_j p_j |\phi_j\rangle\langle \phi_j |
= \rho_{UAB}.
\end{split}
\end{equation}
\begin{equation} \label{convexroof2}
\begin{split}
\tau_{\mathcal A|\mathcal B} = \min &\sum_j p_j (\sigma_{A|UB}(\phi_j) - \sigma_{A|U}(\phi_j)), \\ 
&\sum_j p_j |\phi_j\rangle\langle \phi_j |
= \rho_{UAB}.
\end{split}
\end{equation}
These equations differ from Eqs.~(\ref{aaa}) and (\ref{bbb}) in that we now require the {\em same} decomposition to be used
for both $\sigma_{A|B}$ and $\sigma_{UAB}$ (in Eq.~(\ref{convexroof1})), and for both $\sigma_{A|UB}$ and $\sigma_{A|U}$ 
(in Eq.~(\ref{convexroof2})), whereas Eqs.~(\ref{aaa}) and (\ref{bbb}) entailed separate minimizations.

Let us prove Eq.~(\ref{convexroof1}).  (Eq.~(\ref{convexroof2}) then follows immediately, since for any pure state
$|\phi\rangle$,   $\sigma_{A|UB}(\phi) - \sigma_{A|U}(\phi) = \sigma_{A|B}(\phi) + \sigma_{UAB}(\phi)$.)
First recall that by definition,
\begin{equation}
\tau_{\mathcal A|\mathcal B} = \min \sum_k q_k \tau_{\mathcal A|\mathcal B}(\psi_k), \hspace{1cm} 
 \sum_k q_k |\psi_k\rangle\langle \psi_k | = \rho_{\mathcal A \mathcal B}.
\end{equation}
Let $Q_1 = \{ (q_k, |\psi_k\rangle) \}$ ($Q$ for ``qubit") be a decomposition of $\rho_{\mathcal A \mathcal B}$ that minimizes the average 
of $\tau_{\mathcal A|\mathcal B}(\psi_k)$.
Now to each of these complex pure states $|\psi_k\rangle$, associate two real pure states $|\xi_{k1}\rangle$ and $|\xi_{k2}\rangle$
as in Eq.~(\ref{phi12UAB}).  For each of these pure states, $\sigma_{A|B}(\xi_{ki}) + \sigma_{UAB}(\xi_{ki}) = \tau_{\mathcal A|\mathcal B}(\psi_k)$. 
(This is true because both $\sigma_{A|B}(\xi_{ki})$ and $\sigma_{UAB}(\xi_{ki})$ have the same values for $|\xi_{k1}\rangle$ and
$|\xi_{k2}\rangle$ individually as they have for an equal mixture of $|\xi_{k1}\rangle$ and
$|\xi_{k2}\rangle$.  But this equal mixture is the three-rebit version of $|\psi_k\rangle$, for which we have shown that
$\tau_{\mathcal A|\mathcal B} = \sigma_{A|B} + \sigma_{UAB}$.)
Thus the ensemble of real states $R_1 = \{ (q_1/2, |\xi_{11}\rangle), (q_1/2, |\xi_{12}\rangle), (q_2/2, |\xi_{21}\rangle), (q_2/2, |\xi_{22}\rangle),\ldots \}$ ($R$ for ``rebit'') is a 
decomposition of $\rho_{UAB}$ for which the average value of $\sigma_{A|B}(\xi_{ki})+\sigma_{UAB}(\xi_{ki})$ is equal to $\tau_{\mathcal A|\mathcal B}$.  Therefore the minimum in Eq.~(\ref{convexroof1}) is {\em no larger than}
$\tau_{\mathcal A|\mathcal B}$, since we have found a decomposition that achieves this value.  That is, we have shown
\begin{equation}  \label{ge}
\begin{split}
\tau_{\mathcal A|\mathcal B} \ge  \min &\sum_j p_j (\sigma_{A|B}(\phi_j) + \sigma_{UAB}(\phi_j)), \\
 &\sum_j p_j |\phi_j\rangle\langle \phi_j |
= \rho_{UAB}.
\end{split}
\end{equation}

To show that the inequality also goes in the other direction, let $R_2 = \{ (p_j, |\phi_j\rangle) \}$ be a real decomposition of 
$\rho_{UAB}$ that achieves the minimum in Eq.~(\ref{convexroof1}).  By assumption, $\rho_{UAB}$ is derived from 
the two-qubit state $\rho_{\mathcal A \mathcal B}$, so it is of the form (\ref{rhoUS}), which is invariant under the
transformation $\rho_{UAB} \rightarrow J_U\left( \rho_{UAB} \right)J_U^T$.  (Again we 
are using $J_U$ as shorthand for $J_U \otimes I_A \otimes I_B$.)
Consider, then, the ensemble $R_2' = \{ (p_1/2, |\phi_1\rangle)$, $(p_1/2, J_U|\phi_1\rangle)$, $(p_2/2, |\phi_2\rangle)$, $(p_2/2, J_U|\phi_2\rangle), \ldots \}$, 
which consists of the original states $|\phi_j\rangle$ together with versions of these states in which the ubit has been rotated by $J$.
Since $\rho_{UAB}$ is invariant under such rotations, the new ensemble $R_2'$ is also a decomposition of $\rho_{UAB}$.  Moreover,
the value of $\sigma_{A|B} + \sigma_{UAB}$ is the same for $J_U|\phi_j\rangle$ as it is for $|\phi_j\rangle$.  
Therefore the ensemble $R_2'$ also minimizes the average in Eq.~(\ref{convexroof1}).  Finally,
the two states $|\phi_j\rangle$ and $J_U|\phi_j\rangle$, when mixed together with equal weights, form the rank-2 mixed state
that represents an actual two-qubit pure state $|\psi_j\rangle$, and $\tau_{\mathcal A|\mathcal B}(\psi_j)$ is equal to 
$\sigma_{A|B}(\phi_j) + \sigma_{UAB}(\phi_j)$.  Thus, for the two-qubit ensemble $Q_2 = \{ (p_j, |\psi_j\rangle) \}$, the average 
$\sum_j p_j \tau_{\mathcal A|\mathcal B}(\psi_j)$ is also equal to the minimum value in Eq.~(\ref{convexroof1}).  But 
this ensemble $Q_2$ is a decomposition of $\rho_{\mathcal A \mathcal B}$.  
Therefore, $\tau_{\mathcal A|\mathcal B}(\rho_{\mathcal A \mathcal B})$ (which
is the minimum average $\tau_{\mathcal A|\mathcal B}$ over {\em all} pure-state decompositions of
$\rho_{\mathcal A\mathcal B}$) is no
larger than the minimum value in Eq.~(\ref{convexroof1}).  That is, we have shown
\begin{equation}  \label{le}
\begin{split}
\tau_{\mathcal A|\mathcal B} \le  \min &\sum_j p_j (\sigma_{A|B}(\phi_j) + \sigma_{UAB}(\phi_j)), \\ 
&\sum_j p_j |\phi_j\rangle\langle \phi_j |
= \rho_{UAB}.
\end{split}
\end{equation}
Eqs.~(\ref{ge}) and (\ref{le}) together give us Eq.~(\ref{convexroof1}).   

Note that from Eq.~(\ref{convexroof1}) we can immediately infer the
{\em inequality}
\begin{equation}
\tau_{\mathcal A|\mathcal B} \ge \sigma_{A|B} + \sigma_{UAB}, 
\end{equation}
since the right-hand side allows more freedom than 
Eq.~(\ref{convexroof1}) in the minimization,
by allowing $\sigma_{A|B}$ and $\sigma_{UAB}$ to be evaluated with different decompositions.

Finally, we note that the minimum in Eq.~(\ref{convexroof1}) is well defined even when 
$\rho_{UAB}$ is not of the form (\ref{rhoUS}) and therefore does not correspond to any state of a pair of qubits. 
(The set of three-rebit density matrices is 35-dimensional, whereas the set of two-qubit density matrices
is only 15-dimensional.)  Thus Eq.~(\ref{convexroof1}) gives us a way of extending the concept of the two-qubit tangle
to all three-rebit states.  

\section{Conclusions}

The main focus of this paper has been the rebit version of the three-tangle, which we have defined and evaluated.  We have also
seen how the rebit three-tangle relates to the ordinary tangle between two qubits, when we express the state of the two qubits
as an equivalent three-rebit state: For a pure two-qubit state, the rebit three-tangle is the difference between the two-qubit tangle $\tau_{\mathcal A|\mathcal B}$
and the tangle $\sigma_{A|B}$ associated with the ``rebit shadows'' of $\mathcal A$ and $\mathcal B$.  And for a mixed two-qubit state,
the tangle can be obtained from these rebit quantities by a convex-roof construction.  
Probably the most intriguing result we have seen is that the formula (\ref{realthreetangle}) for the rebit three-tangle is almost identical to the analogous formula (\ref{complexthreetangle})
for the ordinary three-tangle, and yet different in a crucial way (the absence of an overall absolute value sign). 

Of course we have considered here only 
the simplest systems.  Much work has been done in recent years on multipartite entanglement, 
as reviewed in Refs.~\cite{Mintert2, PlenioVirmani, Horodecki, Guhnereview} and as pursued in more recent papers such as 
Refs.~\cite{Ma, Huber} and the papers cited there.
Gour and Wallach, for example, have
identified a relation similar to  
Eq.~(\ref{basictau}) for the case of four qubits \cite{Wallach}.  It would be interesting to know whether the results we have obtained here 
have any simple extensions
to larger systems.     

\section*{Acknowledgement}

This research was supported in part by the Foundational Questions Institute (grant FQXi-RFP3-1350).

\end{document}